\documentclass[onecolumn,showpacs,nobibnotes,nofootinbib,
preprintnumbers]{revtex4}

\usepackage{amsmath,amssymb}
\usepackage{graphicx}

\newcommand{\be}{\begin{equation}}
\newcommand{\ee}{\end{equation}}
\newcommand{\bea}{\begin{eqnarray}}
\newcommand{\eea}{\end{eqnarray}}

\newcommand{\x}{\mathbf{x}}
\newcommand{\y}{\mathbf{y}}

\newcommand{\ab}{\bar\alpha}
\newcommand{\avg}[1]{\left\langle #1 \right\rangle}
\newcommand{\tlam}{\tilde\lambda}
\newcommand{\teps}{\tilde\varepsilon}

\newcommand{\avgrho}{\langle\rho_s\rangle}
\newcommand{\Lr}{\rho}

\newenvironment{caseenum}{

  \begin{enumerate}
}{
  \end{enumerate}

}

\begin{document}

\title{
On the probability distribution of the stochastic saturation scale in QCD}
\author{C. Marquet}
\email{marquet@spht.saclay.cea.fr}
\author{G. Soyez\footnote{on leave from the fundamental theoretical physics 
group of the University of Li\`ege.}}
\email{gsoyez@spht.saclay.cea.fr}
\affiliation{Service de Physique Th\'eorique, CEA/Saclay, 91191 
Gif-Sur-Yvette cedex, France\\URA 2306, unit\'e de recherche associ\'ee au 
CNRS}
\author{Bo-Wen Xiao}
\email{bowen@phys.columbia.edu}
\affiliation{Department of Physics, Columbia University, New York, NY, 10027, 
USA}

\preprint{SPhT-T06/062}
\preprint{CU-TP-1150}

\begin{abstract}
It was recently noticed that high-energy scattering processes in QCD have a 
stochastic nature. An event-by-event scattering amplitude is characterised by 
a saturation scale which is a random variable. The statistical ensemble of 
saturation scales formed with all the events is distributed according to a 
probability law whose cumulants have been recently computed. In this work, we 
obtain the probability distribution from the cumulants. We prove that it 
can be considered as Gaussian over a large domain that we specify and 
our results are confirmed by numerical simulations.
\end{abstract}

\maketitle

\section{Introduction}\label{sec:intro}

The study of the high-energy limit of QCD has recently received important 
contributions coming from analogies with reaction-diffusion systems in 
statistical physics. It has first been shown \cite{mp} that the 
Balitsky-Kovchegov (BK) saturation equation \cite{bk} lies in the same 
universality class as the Fisher-Kolmogorov-Petrovsky-Piscounov (F-KPP) equation 
\cite{fkpp}. The BK equation describes the evolution with rapidity $Y=\log(s)$ 
of the 
dipole scattering amplitude $T(r,Y)$ where $r$ the dipole size. From the analogy 
with the F-KPP equation, one can infer that asymptotic solutions of the BK 
equation are travelling waves. This property states that $T(r,Y)$ is a function 
of the single variable $rQ_s(Y)$ where $Q_s(Y)$ is the {\em saturation scale}.

It has then been realized \cite{MS,fluc} that one also has to take into account 
{\em 
gluon-number fluctuations}. The resulting set of equations can be seen 
\cite{it,ist} as a reaction-diffusion problem. Alternatively, one can write the 
evolution 
equation as a Langevin equation which lies in the same universality class as 
the stochastic F-KPP (sF-KPP) equation 
\cite{fluc}. This is a Langevin equation for an event-by-event scattering 
amplitude which contains a noise term. If one starts with a 
fixed initial condition $T(r,Y_0)$, a single realization of the noise leads 
to an amplitude for a single event, while different realizations of the noise 
result in a dispersion of the solutions.

It has been observed that, event-by-event, the travelling-wave property is 
preserved and that the major effect of this noise term is to introduce 
dispersion in the saturation scale $Q_s$ from one event to another.
The saturation scale becomes thus a random variable. Very recently, the 
cumulants of the probability distribution of the saturation scale (or, more 
precisely, of its logarithm), have been computed \cite{Brunet:2005bz}. In the 
present work, we reconstruct the probability distribution from the cumulants 
and study its different asymptotic behaviours, {\em i.e.} the probability for 
fluctuations
far above, around, or far below the average saturation scale $\bar Q_s$.

We prove that the probability distribution is Gaussian within a large window 
around the average saturation scale. We also compute the probability 
distribution outside of this window and compare our analytical predictions 
with the numerical simulations obtained in \cite{simul}. We justify the use 
of the Gaussian law previously considered in the literature 
\cite{fluc,it,himst,ims} and responsible for a new scaling law of the dipole 
amplitude: when computing the physical amplitude by averaging over all events 
using a Gaussian distribution, one gets $T(r,Y)=T(\log[r^2\bar 
Q_s^2(Y)]/\sqrt{Y})$.

The structure of the paper is as follows. We start in section \ref{sec:qcd} 
by introducing the QCD evolution equations and their link with the sF-KPP 
equation. In section \ref{sec:prob}, we compute the probability distribution 
from the cumulants obtained in \cite{Brunet:2005bz} and compare those 
predictions with numerical simulations. Finally, we discuss the implications 
on physical amplitudes in section \ref{sec:amp}. We conclude in section 
\ref{sec:ccl}.

\section{Stochasticity in high-energy QCD evolution}\label{sec:qcd}

\begin{figure}
\includegraphics{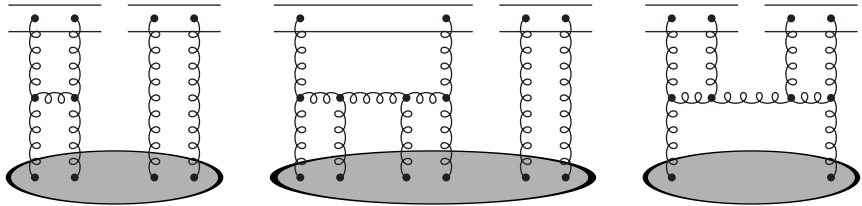}
\caption{These diagrams show the contributions to the evolution of $T^{(2)}$, 
the scattering amplitude for a set of two dipoles off an arbitrary target 
represented here by a gray disk. From left to right: 
a linear BFKL term proportional to $T^{(2)}$, a merging term proportional
to $T^{(3)}$ responsible for saturation, and a splitting term proportional to 
$T$ responsible for fluctuations.}
\label{fig:t2}
\end{figure}

To begin with, let us recall the important points concerning the QCD evolution 
equations towards high energy. To the present knowledge (in the leading 
logarithmic approximation and in the large-$N_c$ limit), those equations give 
the energy evolution for the scattering amplitude $\avg{T^{(k)}}$ between a 
system of $k$ dipoles of transverse coordinates 
$(\x_1,\y_1),\dots,(\x_k,\y_k)$ and a generic target. They include 
three different types of contributions \cite{it} as shown in Fig. 
\ref{fig:t2}: BFKL ladders, gluon mergings responsible for saturation 
corrections, and gluon splitting corresponding to gluon-number fluctuations.  
Summing these contributions results in a infinite hierarchy of equations, 
containing Pomeron loops, in which the evolution of $\avg{T^{(k)}}$ depends 
on $\avg{T^{(k)}}$, $\avg{T^{(k+1)}}$ and $\avg{T^{(k-1)}}$, coming 
respectively from the three types of contributions discussed previously. 

The solutions of this hierarchy are not yet known and it is useful to deal 
with a simplified version of it. If we integrate out the impact-parameter 
dependence by performing a coarse-graining approximation \cite{it}, we can 
Fourier 
transform the set of equations to momentum space ($r\rightarrow k$). The 
resulting hierarchy becomes equivalent to the following Langevin equation
\begin{equation}\label{eq:langevin}
\partial_Y T(k,Y) = \ab\int_0^\infty \frac{dp^2}{p^2}\,
\left[
\frac{p^2 T(p,Y)-k^2T(k,Y)}{\left| p^2-k^2 \right|}+\frac{k^2 
T(k,Y)}{\sqrt{4p^4+k^4}}
\right]
-\ab T^2(k,Y) + \ab \sqrt{2\kappa\alpha_s^2T(k,Y)}\:\nu(k,Y),
\end{equation}
where $\nu(k,Y)$ is a Gaussian white noise satisfying
\be
\avg{\nu(k,Y)} = 0\qquad\text{ and }\quad
\avg{\nu(k,Y)\nu(k',Y')} = 
\frac{2}{\ab\pi}\delta\left(\log(k^2/k'^2)\right)\delta(Y-Y').
\ee
The factor $\kappa$ present in front of the noise term comes from a 
local-noise approximation performed during the coarse graining.

\begin{itemize}
\item The linear part of equation \eqref{eq:langevin} is the BFKL equation 
\cite{bfkl}. 
Its solution $T_{\text{lin}}(k,Y)$ is a superposition of waves:
\be T_{\text{lin}}(k,Y) = 
\int_{\frac12-i\infty}^{\frac12+i\infty}\frac{d\gamma}{2i\pi}\,
e^{-\gamma[L-\bar\alpha v(\gamma)Y]}\,T_0(\gamma)
\ee 
where $L=\log(k^2/k_0^2)$ and $T_0(\gamma)$ specifies the initial condition, 
characterised by the 
reference scale $k_0$. Each wave has a different speed $v(\gamma)$ given by 
\be\label{eq:bfkl_kernel}
v(\gamma)=\frac{\chi(\gamma)}{\gamma} \qquad \mbox{ with } \qquad 
\chi(\gamma)=2\psi(1)-\psi(\gamma)-\psi(1-\gamma)\ ,\ee
$\psi(\gamma)=\frac{d}{d\gamma}\log[\Gamma(\gamma)]$ being the 
digamma function. 

\item The equation \eqref{eq:langevin} without the noise term is the BK 
equation. It has been shown \cite{mp} to lie in the same universality class as 
the F-KPP equation \cite{fkpp} which allowed analytical predictions. The 
non-linear term damps the growth of the amplitude predicted by the linear BFKL
equation in such a way that the asymptotic solution $T_{\text{tw}}(k,Y)$ of the 
BK equation is a travelling wave of minimal speed $T_{\text{tw}}(L-\ab v_c Y).$ 
The minimal 
speed $v_c$ is obtained for a value of $\gamma$ that we shall denote $\gamma_c:$
\be
v_c=\min_\gamma \,v(\gamma)\equiv v(\gamma_c)\quad \mbox{ with } \gamma_c
\mbox{ solution of }\quad \frac{\chi(\gamma)}{\gamma}=\chi'(\gamma)\ .
\ee
For the BFKL kernel \eqref{eq:bfkl_kernel}, one has $\gamma_c=0.6275$ and 
$v_c=4.883.$  From the position of the 
travelling wave one obtains the saturation scale $Q_s(Y)$, the momentum for 
which $T_{\text{tw}}$ is constant:
\be\label{eq:trav}
T_{\text{tw}}(L-\ab v_c Y) \stackrel{L\gg \ab v_c Y}{=} 
e^{-\gamma_c(L-\bar\alpha v_c Y)}=
\left(\frac{k^2}{Q_s^2(Y)}\right)^{-\gamma_c} \mbox{ with } 
Q_s^2(Y)=k_0^2\ e^{\bar\alpha v_c Y}\ .\ee

\item The complete equation is a stochastic equation: it is the BK equation 
supplemented with a noise term which accounts for gluon-number fluctuations. 
It has been shown that, in the diffusive approximation where the 
integration in \eqref{eq:langevin} is expressed as a second-order differential 
operator, Eq. \eqref{eq:langevin} lies \cite{fluc} in the same universality 
class as the sF-KPP equation
\begin{equation}\label{eq:sfkpp}
\partial_t u(x,t) = \partial_x^2 u(x,t) + u(x,t) - u^2(x,t) + \sqrt{2\lambda 
u(x,t) [1-u(x,t)]} \, \eta(x,t),
\end{equation}
with
\be
\avg{\eta(x,t)} = 0\qquad\text{ and }\quad
\avg{\eta(x,t)\eta(x',t')} = \delta(t-t')\delta(x-x').
\ee
In the analogy between \eqref{eq:langevin} and \eqref{eq:sfkpp}, $\ab Y$ 
plays the role of time $t$ while the space variable $x$ is related to $L$ and 
$\lambda\propto \kappa\alpha_s^2$. The main analytical predictions concerning 
\eqref{eq:langevin} are obtained from the study of the sF-KPP equation in the 
weak-noise limit. First, if one compares solutions of \eqref{eq:langevin} 
with solutions of the BK equation, one obtains that each individual event is 
a travelling wave with a speed smaller than the one predicted from the BK 
equation. Then, if one consider a whole set of events, the main effect of the 
noise is to introduce dispersion in the position of different events. 
Physically, this means that the saturation scale fluctuates from one event to 
another and we can show, analytically (in the weak or strong noise limit) and 
numerically, that the average saturation scale $\bar Q_s^2(Y)$ grows as 
$\exp(\ab v Y)$ with $v<v_c$
and that the dispersion of 
$\log(Q_s^2(Y))$ 
around that average value increases like $\sqrt{Y}$ as expected from a 
random-walk process. 
\end{itemize}

\section{The probability distribution of the stochastic saturation 
scale}\label{sec:prob}

The saturation scale being a stochastic variable, it is characterised by a 
probability distribution. Let us call $P(\rho_s)$ the probability 
distribution for the variable $\rho_s=\log(Q_s^2(Y)/k_0^2).$ From general 
arguments it was first inferred that $P(\rho_s)$ should be Gaussian with an 
average value $\langle\rho_s\rangle=\log(\bar{Q}_s^2(Y)/k_0^2)=\bar\alpha v Y$ 
and a 
variance $\sigma^2=\bar\alpha D Y,$ 
with $v$ and $D$ are coefficients characterising the speed and dispersion 
of the travelling wavefronts. This was then used extensively in the literature 
\cite{fluc,it,himst,ims}. Recently, 
the cumulants $\kappa_n$ of the distribution have been computed. The first 
cumulant $\kappa_1=\langle\rho_s\rangle$ is the average value of $\rho_s$ and 
the second cumulant $\kappa_2=\sigma^2$ is the variance of the distribution. In 
the Gaussian case higher-order cumulants are zero. The calculation 
of~\cite{Brunet:2005bz} showed that this was not the case: they find that 
higher-order cumulants are proportional to the second cumulant. To summarise, 
one has
\be\label{eq:cumul}
\kappa_1=\langle\rho_s\rangle=\bar\alpha v Y\ ,\hspace{1cm}
\kappa_2=\sigma^2=\bar\alpha D Y\ ,\hspace{1cm}
\kappa_n=\frac{3\gamma_c^2}{\pi^2}\frac{n!\zeta(n)}{\gamma_c^n}\sigma^2\ ,
\ee
where $\zeta(n)$ is the Riemann Zeta function. Note that these results are valid 
in the high-energy limit $\gamma_c\sigma^2\gg 1$.

In the weak-noise ($\alpha_s^2 \ll 1$) limit in which the results were obtained, 
one has
\be
v = v(\gamma_c)
    - \frac{\pi^2\gamma_c^2v''(\gamma_c)}{\log^2(1/\alpha_s^2)}
    \left(1-\frac{3\log[\log(1/\alpha_s^2)]}{\log(1/\alpha_s^2)}\right)
\qquad\text{ and }\quad
D=\frac{\pi^4\gamma_c v''(\gamma_c)}{3\log^3(1/\alpha_s^2)}\ .
\ee
But the fact that $\langle\rho_s\rangle$ and $\sigma^2$ are proportional to 
$Y$ is believed to be more general. This has been shown by an analytical 
study of the strong noise limit \cite{strong} and it is confirmed by numerical 
simulations \cite{simul,simul2} for arbitrary values of the noise strength (see 
also \cite{simul3} for related numerical studies). One 
observes that, when the noise strength increases, the speed of the wave $v$ 
decreases and the dispersion coefficient $D$ increases. In what follows, we 
shall therefore keep $v$ and $D$ as parameters.

\subsection{Analytical results}

In this section, we compute the probability distribution $P(\rho_s)$ from the 
cumulants. Our starting point is the generating function for the moments of 
$P(\rho_s)$
\begin{equation}\label{eq:genp}
\left\langle e^{\lambda\rho_s}\right\rangle = \int_{-\infty}^\infty 
d\rho_s\,e^{\lambda \rho_s}\,P(\rho_s)\ .
\end{equation}
The cumulants generating function then reads
\begin{equation}\label{eq:cumul_sum}
\log \left\langle e^{\lambda \rho_s}\right\rangle 
=\sum_{n>0}\frac{\kappa_n\lambda^n}{n!}
 = \avgrho \lambda + \frac{3\gamma_c^2}{\pi^2}\sigma^2 \sum_{n=2}^{\infty 
}\frac{\zeta(n)\lambda^n}{\gamma_c^n}\ .
\end{equation}
If we use the following integral representation of the Zeta function
\be
\zeta(n) = \frac{1}{\Gamma(n)} \int_0^\infty du\,\frac{u^{n-1}}{e^u-1},
\ee
we are able to compute analytically the cumulants generating function 
\eqref{eq:cumul_sum}
\be
\log \left\langle e^{\lambda \rho_s}\right\rangle 
  =  \avgrho \lambda + \frac{3\gamma_c^2}{\pi^2}\sigma^2 
\frac{\lambda}{\gamma_c} \int_0^\infty 
du\,\frac{e^{\lambda u/\gamma_c}-1}{e^u-1}
  =  \avgrho \lambda - \frac{3\gamma_c\sigma^2}{\pi^2}\lambda \left[\gamma_E + 
\psi\left(1-\frac{\lambda}{\gamma_c}\right)\right] ,
\ee
where $\gamma_E \approx 0.577216$ is the Euler constant.We can then inverse the 
Laplace 
transform in \eqref{eq:genp} to obtain the probability distribution ($c < 
\gamma_c$)
\begin{equation}\label{eq:proba_full}
P(\rho_s) = \int_{c-i\infty}^{c+i\infty} \frac{d\lambda}{2i\pi}\,
            \exp\left\{-\lambda z
                       - b\lambda \left[\gamma_E + 
\psi\left(1-\frac{\lambda}{\gamma_c}\right) \right] 
                \right\} \ .\end{equation}
where we have introduced
\be 
z=\rho_s-\langle\rho_s\rangle=\log\left(\frac{Q_s^2}{\bar{Q}_s^2}\right)\ ,
\hspace{1cm}b=\frac{3\gamma_c\sigma^2}{\pi^2}\ .
\ee
$z$ is the distance of the saturation scale to its average value, and $b$ is a 
convenient redefinition of the variance. To evaluate further the probability 
distribution $P(\rho_s)$, we shall perform the integration over $\lambda$ in the 
saddle point approximation. One finds that the saddle point $\tlam\equiv 
\gamma_c(1-\teps)$ has to satisfy
\begin{equation}\label{eq:saddle}
z + b\left[\gamma_E + \psi(\teps) - (1-\teps)\psi^{(1)}(\teps)\right] =0\ ,  
\end{equation}
where $\psi^{(n)}(x)$ is the polygamma function defined 
as $\frac{d^n}{dx^n}\psi(x)$. The probability distribution is then given by
\begin{equation}\label{eq:proba}
P(\rho_s) = 
\left\{\frac{12\sigma^2}{\pi}\left[\psi^{(1)}(\teps)-\frac{1-\teps}{2}
\psi^{(2)}(\teps)\right]\right\}^{-1/2}
            \exp\left[-\frac{3\sigma^2}{\pi^2} \gamma_c^2(1-\teps)^2 
\psi^{(1)}(\teps)\right].
\end{equation} Although Eq.~\eqref{eq:saddle} has no exact analytical 
solution, one can solve it in three interesting limits: 
$(i)$ $z/b \rightarrow 0$ which corresponds to $\teps \rightarrow 1$; 
$(ii)$ $z/b\rightarrow \infty $ related to the limit $\teps\rightarrow 0$; 
and $(iii)$ $z/b \rightarrow -\infty $ equivalent to the limit 
$\teps\rightarrow \infty$. We detail these limits hereafter.

\begin{caseenum}

\item {\bf When $z/b \rightarrow 0$ or equivalently $|z|\ll 
\gamma_c\sigma^2$}. Eq.~\eqref{eq:saddle} becomes
\be
\frac{z}{b} + \left[ \gamma_E+\psi(1)-2(1-\teps)\psi^{(1)}(1) \right] = 0 
\qquad\Rightarrow\qquad \teps=1-\frac{3z}{\pi^2 b}\ ,
\ee
where we have used $\psi(1)=-\gamma_E$ and $\psi^{(1)}(1)=\pi^2/6$. After 
replacement in \eqref{eq:proba} we get
\begin{equation}\label{eq:proba_gauss_full}
P(\rho_s) \approx \frac{1}{\sqrt{2\pi 
\sigma^2}}\left[1-\frac{9\zeta(3)}{\pi^2}\frac{z}{\gamma_c\sigma^2}\right]
\exp\left(-\frac{z^2}{2\sigma^2}\right).
\end{equation}
We kept the first subleading term to show that this distribution has a maximum 
for the constant value
\begin{equation}\label{eq:shift}
z \approx \frac{-9\zeta(3)}{\pi^2\gamma_c} \approx -1.75.
\end{equation}
The most probable value for the saturation scale is therefore not the average 
value $\avgrho$.
For asymptotically large energies, we recover
\begin{equation}\label{eq:proba_gauss}
P(\rho_s) \approx \frac{1}{\sqrt{2\pi \sigma^2}} 
\exp\left(-\frac{z^2}{2\sigma^2}\right)
\end{equation}
The probability distribution of fronts is Gaussian around the average. We 
shall 
discuss later on the fact that this is the dominant behaviour at high-energy 
which justifies the use of a Gaussian distribution in \cite{fluc,it,himst,ims}. 

\item {\bf When $z/b\rightarrow +\infty$ or equivalently $z\gg 
\gamma_c\sigma^2$}. We are sensitive to small values of $\teps$. One can thus 
simplify Eq.~\eqref{eq:saddle} and obtain
\begin{equation}
-\left( \frac{z}{b}\right) + 
\frac{1}{\teps}+\frac{1-\teps}{\teps^2}+\frac{\pi^2}{6}=0\ ,
\qquad\Rightarrow\qquad 
\teps=\sqrt{\frac{3}{\pi^2}\frac{\gamma_c\sigma^2}{z}}
\left(1+\frac{\gamma_c\sigma^2}{4z}\right).
\end{equation}
After a bit of algebra, this turns into the following limit for the 
probability distribution
\begin{equation}\label{eq:proba_large}
P(\rho_s) \approx 
\frac{3^{1/4}}{2\pi\sigma}\left(\frac{\gamma_c\sigma^2}{z}\right)^{3/4}
\exp\left[ 
-\gamma_c z \left(1-2\sqrt{\frac{3}{\pi^2}\frac{\gamma_c\sigma^2}{z}} \right) 
\right]\ .
\end{equation}
This corresponds to a power-law tail $P(\rho_s)\sim (Q_s/\bar Q_s)^{-2\gamma_c}$ 
at very large $Q_s$.

\item {\bf When $z/b\rightarrow -\infty $ or equivalently $z\ll 
-\gamma_c\sigma^2$}. We have $\teps\to \infty$, and Eq.~\eqref{eq:saddle} 
turns into
\begin{equation}
\frac{z}{b}+\gamma_{E}+\log(\teps)+1-\frac{1}{\teps}=0
\qquad\Rightarrow\qquad \teps = 
\exp\left(-\frac{\pi^2}{3}\frac{z}{\gamma_c\sigma^2}-1-\gamma_E\right)+1.
\end{equation}
The probability \eqref{eq:proba} is then found to be
\begin{equation}\label{eq:proba_small}
P(\rho_s) \approx \sqrt{\frac{\pi}{6}}\frac{1}{\sigma}
 \exp\left\{-\frac{\pi^2}{6}\frac{z}{\gamma_c\sigma^2}
           -\frac{1+\gamma_E}{2}-\frac{3\gamma_c^2\sigma^2}{\pi^2}\left[
	   \exp\left(-\frac{\pi^2}{3}\frac{z}{\gamma_c\sigma^2}-1-\gamma_E
               \right)-\frac{1}{2}  \right] \right\}.
\end{equation}
This is a Gumbel distribution, which goes to zero very fast when  $z \ll 
-\gamma_c\sigma^2$.

\end{caseenum}

As a function of $z$ starting from $z=-\infty,$ the transition between the 
regime \eqref{eq:proba_small} and the regime \eqref{eq:proba_gauss} happens 
for $z=-\gamma_c \sigma^2$ and the transition between the regime 
\eqref{eq:proba_gauss} and the regime \eqref{eq:proba_large} happens for 
$z=\gamma_c \sigma^2.$ For those two points, 
$\rho_s=\langle\rho_s\rangle\pm\gamma_c \sigma^2,$ the probability is of 
order $e^{-\gamma_c^2\sigma^2}/\sigma$ which is very small. This means that 
the probability $P(\rho_s)$ is not Gaussian only for very improbable 
fluctuations. To describe the bulk of the events, the Gaussian distribution 
\eqref{eq:proba_gauss} is a good approximation.

\subsection{Numerical results}

In this section, we shall compare the analytical prediction \eqref{eq:proba} 
derived in the previous section with the numerical simulation of 
\eqref{eq:langevin} introduced in \cite{simul}. 
To fix things properly we need first to recall a few points from 
\cite{simul}. 
In that study, we start with a fixed initial condition and evolve numerically 
the QCD Langevin equation with different realisation of the noise 
term (the precise method used to solve the equation is not important 
for our purposes here, so we refer to \cite{simul} for details). The momentum 
$k$ is discretised in bins of $L=\log(k^2/k_0^2)$ between $L_{\text{min}}$ and 
$L_{\text{max}}$. Hence, the numerical simulations of \eqref{eq:langevin} 
results in a set of events $T_i(L_j,Y)$, with $i=1,\dots,n_{\text{ev}}$, {\em 
i.e.}, for each event $i$, we get the rapidity evolution of the amplitude in 
each momentum bin.

As expected from the analytical study of \eqref{eq:langevin} in the 
weak-noise limit, the numerical studies have confirmed that each event
is a travelling wave whose speed decreases when the strength of the fluctuations 
increases. Also, at large rapidities, the position of the wavefront shows a 
dispersion proportional to $\sqrt{Y}$ if one consider a whole set of events.
Practically, those results can be summarised by saying that, for the $i$th 
event, we have
\be
T_i(L,Y) = T_i(L-L_{s,i}(Y))\qquad \text{ with }\quad L_s \equiv \rho_s = 
\log(Q_s^2(Y)/k_0^2).
\ee

The position $L_{s,i}(Y)$ of the wavefront for one event can be extracted 
from 
the numerical simulations by solving $T_i(L,Y)=T_0$ for a fixed $T_0$ (here, 
we adopted $T_0=0.2$) and for different values of the rapidity. By measuring 
(the logarithm of) the saturation scale, one can obtain its statistical 
distribution. More precisely, for each rapidity, we construct a histogram by 
counting, for a set of events, the number of event for which the saturation 
scale is in each momentum bin.

In order to compare with the analytical predictions, we shall, instead of the 
histogram for $\rho_s$, use the distribution for $z=\rho_s-\avgrho$. 
Normalised in such a way that its integral is 1, it is directly comparable 
with \eqref{eq:proba} if one fixes the dispersion parameter. For the latter 
we adopt the value computed numerically from the histogram of $\rho_s$ or 
$z$. In other words, we fix the two lowest-order cumulants 
$\avgrho$ 
and $\sigma^2$, and compare the resulting analytical and numerical probability 
distribution at different rapidities.

\begin{figure}
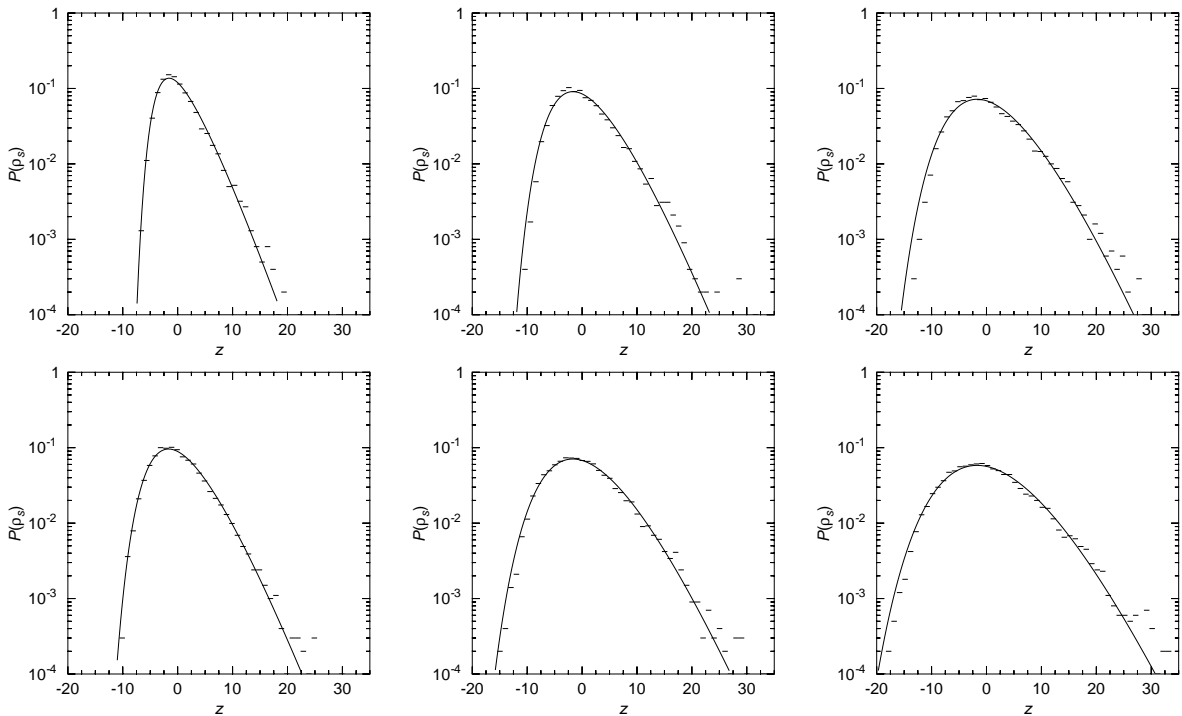

\includegraphics[scale=0.5]{fig-0.200-20.ps}\hspace{0.4cm}
\includegraphics[scale=0.5]{fig-0.200-30.ps}\hspace{0.4cm}
\includegraphics[scale=0.5]{fig-0.200-40.ps}\\
\includegraphics[scale=0.5]{fig-1.000-20.ps}\hspace{0.4cm}
\includegraphics[scale=0.5]{fig-1.000-30.ps}\hspace{0.4cm}
\includegraphics[scale=0.5]{fig-1.000-40.ps}
\caption{Comparison of the probability distribution \eqref{eq:proba} as a 
function of $z=\rho_s-\avgrho$ with results from numerical simulations with 
$\tilde\kappa=1$ (top) and $\tilde\kappa=5$ (bottom). The comparisons are 
made for $Y=20$, $30$, $40$, from left to right.}\label{fig:kappa}
\end{figure}

The results are shown in Fig. \ref{fig:kappa}, for two 
sets of 10000 events taken from \cite{simul}. Those two sets corresponds to 
different values of the noise strength 
$\tilde\kappa\!=\!10\pi\kappa/N_c^2\!=\!1$ and $5$, the value of $\ab$ being 
fixed to 0.2. For 
both cases, we show the probability distribution for rapidities $Y=20$, $30$ and 
$40$. Those rapidities are in the region where the travelling wavefront has 
reached its asymptotic behaviour and the dispersion of the wavefronts is 
proportional to $\sqrt{Y}$ as expected. 

It is obvious from Fig. \ref{fig:kappa} that the agreement is excellent with 
predictions from \eqref{eq:proba}. One can see that, in the bulk of the 
distribution, the probability is Gaussian, up to a shift of the maximum to a 
negative value which is consistent with \eqref{eq:shift}. We also observe that 
the probability falls very fast in the infrared and has a tail which favours 
fluctuations to large values of the saturation momentum. However, the deviations 
from a Gaussian behaviour only appear for events with a very small probability.

\section{Consequences for the physical scattering amplitude}
\label{sec:amp}

We shall now compute the behaviour of the average scattering amplitudes 
$\avg{T(\Lr,Y)}$ where $\rho = \log[1/(k_0^2r^2)]$, using the probability 
distribution obtained in the previous section. For simplicity, we consider the 
amplitude in coordinate space.
The average amplitude can be directly computed from the probability $P(\rho_s)$ 
using
\begin{equation}\label{eq:average}
\avg{T(\Lr,Y)} = \int_{-\infty}^\infty d\rho_s\,P(\rho_s)T(\Lr-\rho_s),
\end{equation}
where $T(\Lr-\rho_s)$ describe the event-by-event amplitude which features the 
travelling-wave property w.r.t. the saturation scale $\rho_s$. 
Inserting the probability distribution \eqref{eq:proba_full} into 
\eqref{eq:average} and using the Mellin representation
\be
T(\Lr) = \int \frac{d\lambda}{2i\pi}\,e^{-\lambda \Lr}\,\tilde T(\lambda), 
\ee
we obtain for the physical amplitude
\begin{equation}\label{eq:ampl_full}
\avg{T(\Lr,Y)} = \int \frac{d\lambda}{2i\pi}\, \tilde T(\lambda)\, 
\exp\left\{-\lambda Z
            - b\lambda \left[\gamma_E + 
\psi\left(1-\frac{\lambda}{\gamma_c}\right) \right] 
            \right\}.
\end{equation}
We have introduced the convenient variable
\be Z=\Lr-\avgrho=-\log\left(r^2\bar{Q}_s^2(Y)\right)\ .\ee

Since the function $\tilde T(\lambda)$ is slowly varying the remaining 
integration will basically be sensitive to the same saddle point 
\eqref{eq:saddle} as for the computation of the probability. The main difference 
lies in the fact that, because of the unitarity constraint, $\tilde T(\lambda)$ 
has a pole in $\gamma=0$.
Hence, in order to get relevant expressions in the different physical limits, 
the following assumption for the event-by-event wavefront is sufficient:
\begin{equation}\label{eq:front}
T(\Lr-\rho_s) = \begin{cases} 
1                               & \Lr \le \rho_s\\
\exp\left[-\gamma_0(\Lr-\rho_s)\right] & \Lr > \rho_s
\end{cases} \qquad\qquad\Rightarrow\qquad \tilde T(\lambda)= 
\frac{1}{\lambda}+\frac{1}{\gamma_0-\lambda}.
\end{equation}
This satisfies both the travelling-wave property \eqref{eq:trav} and unitarity 
requirements. The exponent $\gamma_0$ can be chosen\footnote{Although we keep it 
as a variable, one should probably adopt $\gamma_0=\gamma_c$. This exponent in 
the tail differs from the usual $\gamma_0=1,$ however, since the 
results \eqref{eq:cumul} are obtained in the weak-noise limit, the tail of the 
wavefront extends far above the saturation momentum and the proper matching 
with the $\exp(-\Lr)$ behaviour should only lead to higher-order corrections.} 
between $\gamma_c$ and 1.

We can now introduce \eqref{eq:front} in \eqref{eq:ampl_full} and we get
\begin{equation}\label{eq:average_full}
\avg{T(\Lr,Y)} = \int_{c-i\infty}^{c+i\infty} 
\frac{d\lambda}{2i\pi}\,\frac{\gamma_0}{\lambda(\gamma_0-\lambda)}
            \exp\left\{-\lambda Z 
            - b\lambda \left[\gamma_E + 
\psi\left(1-\frac{\lambda}{\gamma_c}\right) \right] 
            \right\},
\end{equation}
where the contour of integration is now restricted to values of $c$ such that 
$0<c<\gamma_c$. The remaining integration over $\lambda$ can be computed in the 
saddle-point approximation and the condition \eqref{eq:saddle} is still 
appropriate for \eqref{eq:average_full}.
Again, the physical picture arises when we consider the three different limits 
$|Z|\ll\gamma_c\sigma^2$, $Z\gg \gamma_c\sigma^2$ and $Z\ll -\gamma_c\sigma^2$, 
which we detail hereafter.
\begin{caseenum}
\item {\bf Case} $Z\gg \gamma_c\sigma^2$: as for the computation of the 
probability, the saddle point is obtained by expansion around $\teps\to 0$ or 
$\tlam \to \gamma_c$ and the same result is obtained, up to an extra factor 
$(\gamma_0-\tlam)^{-1}$. For $\gamma_0>\gamma_c$, this leads to
\begin{equation}\label{eq:average_large0}
\avg{T(\Lr,Y)} \approx 
\frac{3^{1/4}}{2\pi\sigma(\gamma_0-\gamma_c)}\left(\frac{\gamma_c\sigma^2}{Z}
\right)^{3/4}\exp(-\gamma_c Z ).
\end{equation}
If $\gamma_0=\gamma_c$, we get an additional contribution from the 
pole\footnote{If we treat properly the pole inside the saddle-point equation, it 
only introduces a factor $(\gamma_c-\tlam)^{-1}$ which is subleading compared to 
the $(\gamma_c-\tlam)^{-2}$ from the $\psi^{(1)}$ function. Hence the saddle 
point remains unchanged.}
\begin{equation}\label{eq:average_large}
\avg{T(\Lr,Y)} \approx 
\frac{3^{-1/4}}{2\gamma_c\sigma}\left(\frac{\gamma_c\sigma^2}{Z}\right)^{1/4}
\exp(-\gamma_c Z).
\end{equation}
\item {\bf Case} $Z\ll -\gamma_c\sigma^2$: this case is slightly more intricate. 
Indeed, since $\tlam\to-\infty$, the contour of integration must first be moved 
towards $c=\tlam$. The integration therefore receives a contribution of $1$ 
coming from the pole at $\lambda=0$. The final result reads
\begin{equation}\label{eq:average_small}
\avg{T(\Lr,Y)} \approx 1-\sqrt{\frac{\pi}{6}}\frac{1}{\gamma_c\sigma}
                      \exp\left\{\frac{\pi^2}{2}\frac{Z}{\gamma_c\sigma^2}       
                +\frac{3}{2}(1+\gamma_E)-\frac{3\gamma_c^2\sigma^2}{\pi^2}
	\left[                       
\exp\left(-\frac{\pi^2}{3}\frac{Z}{\gamma_c\sigma^2}-1-\gamma_E
			       \right)-\frac12\right]
			  \right\}.
\end{equation}
\item {\bf Case} $|Z|\ll\gamma_c\sigma^2$: for that last case, we have to 
distinguish between two possibilities: $Z>0$ and $Z<0$. Let us start with the 
positive values of $Z$: as for the computation of the probability, this 
situation is sensitive to $\tlam \to 0$. Hence, we need to properly take into 
account the prefactor $\lambda^{-1}=\exp[-\log(\lambda)]$. This gives an 
additional contribution $-\tlam^{-1}$ to the saddle point equation which becomes
\be
Z - \sigma^2 \tlam - \frac{1}{\tlam} = 0 \qquad\Rightarrow\qquad \tlam = 
\frac{Z}{2\sigma^2} + \frac{1}{2\sigma^2} \sqrt{Z^2+4\sigma^2}.
\ee
If, in addition, one requires $Z\gg \sigma$, we recover $\tlam=Z/\sigma^2$ as 
for the computation of the probability. The average amplitude is then the same 
as the probability, up to an extra factor $\lambda^{-1}$, which gives
\begin{equation}\label{eq:average_gauss_pos}
\avg{T(\Lr,Y)} \approx 
\frac{1}{\sqrt{2\pi}}\frac{\sigma}{Z}\exp\left(-\frac{Z^2}{2\sigma^2}\right).
\end{equation}
For the case where $Z$ is negative, we first move the integration contour on the 
left side of the pole at $\lambda=0$ which gives a contribution of $1$ and the 
remaining integration is computed in the same way as for positive $Z$. That is, 
for $-\gamma_c\sigma^2 \ll Z \ll -\sigma$, we obtain
\begin{equation}\label{eq:average_gauss_neg}
\avg{T(\Lr,Y)} \approx 1 - 
\frac{1}{\sqrt{2\pi}}\frac{\sigma}{|Z|}\exp\left(-\frac{Z^2}{2\sigma^2}\right),
\end{equation}
where we have explicitly emphasised the sign of the second term.
The two results \eqref{eq:average_gauss_pos} and \eqref{eq:average_gauss_neg} 
can be summarised in only one formula:
\begin{equation}\label{eq:average_gauss}
\avg{T(\Lr,Y)} 
\approx \frac{1}{2}\,{\rm erfc}\left(\frac{Z}{\sqrt{2}\sigma}\right),
\end{equation}
where ${\rm erfc}(x)$ is the complementary error function.
\end{caseenum}

\begin{figure}
\includegraphics[width=12cm]{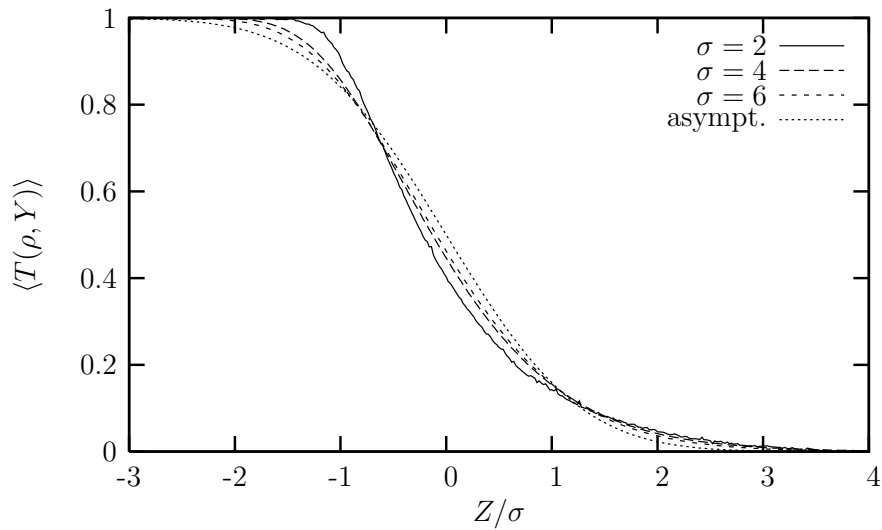}
\caption{Comparison of the physical amplitude $\avg{T(\Lr,Y)}$ plotted as a 
function of $Z/\sigma$ and the error function obtained as the asymptotic 
behaviour.}\label{fig:tcomp}
\end{figure}

The results \eqref{eq:average_large}, $\eqref{eq:average_small}$ and 
\eqref{eq:average_gauss} call for a number of comments. The most important one 
is that (for very high energies such that $\gamma_c\sigma^2\gg 1$) one recovers 
the error function as expected previously in the literature 
\cite{fluc,it,himst,ims} and the scattering amplitude satisfies {\em diffusive 
scaling} as it depends only on the ratio $Z/\sigma$. This result is valid in a 
very large window around the average saturation scale ($Z\ll \gamma_c 
\sigma^2$). Since this can directly be derived from the front \eqref{eq:front} 
convoluted with the Gaussian probability distribution \eqref{eq:proba_gauss}, 
this means that it is sufficient to consider a Gaussian distribution in order to 
compute the scattering amplitudes at high-energy. In addition, the error 
function fully comes from $\Lr<\rho_s$ in \eqref{eq:front}, which proves the 
expected result \cite{himst,ims} that the amplitude is dominated, up to very 
large values of $\Lr$, by fronts at saturation {\em i.e.} by {\em black spots}. 
Finally, let us notice that our argument does not depend on the particular 
choice for the event-by-event front. Indeed, the pole in $\lambda^{-1}$ comes 
from the saturated part of \eqref{eq:front} and thus the condition that the 
event-by-event amplitude satisfies unitarity is sufficient to obtain 
\eqref{eq:average_gauss}.

To illustrate that behaviour, we have computed from \eqref{eq:average} the 
amplitude obtained using an event-by-event front $\Theta(\rho_s-\Lr)$ 
($\Theta(x)$ being the Heaviside function). We plot on Fig. \ref{fig:tcomp} the 
result for different values of $\sigma$ as a function of the high-energy scaling 
variable $Z/\sigma$. We clearly observe that when $\sigma$ increases ({\em i.e.} 
when energy increases), we converge to the universal error function 
\eqref{eq:average_gauss}.

The behaviour in the far tail $Z\gg\gamma_c \sigma^2$ is also quite interesting. 
Indeed, if we adopt $\gamma_0=\gamma_c$, the amplitude \eqref{eq:average_large} 
decreases as $\exp(-\gamma_c Z)$ as expected, but it receives contributions both 
from the event-by-event amplitude and from the probability distribution. 
Finally, deep inside the saturation region we obtain that the amplitude reaches 
the unitarity limit like $1-c_1\exp[-c_2(r\bar Q_s)^{2c_3}]$ where $c_1$, $c_2$ 
and $c_3$ can be read on Eq. \eqref{eq:average_small}.

\section{Conclusion and discussion}\label{sec:ccl}

Let us now summarise the results obtained in this paper. First, we have shown 
that 
it 
is possible to compute the probability distribution for the position of the 
wavefront satisfying the stochastic F-KPP equation, which corresponds to the 
saturation scale in QCD. Our starting point is the cumulants of this 
distribution computed previously in \cite{Brunet:2005bz}. The probability 
distribution is computed in a saddle point approximation. This allows us to 
compute 
its leading behaviour in the vicinity of the average saturation scale $\avgrho$ 
where the probability is Gaussian, as well as for improbable events far above or 
below $\avgrho$.

Then, we have checked that the predictions for the probability distribution 
are in good agreement with the results obtained from the numerical 
simulations of the QCD Langevin equation \eqref{eq:langevin}, once one fixes the 
average 
position and its dispersion from their numerical values. Since the numerical 
simulations are not obtained in the weak noise limit and not directly for the 
sF-KPP equation, this proves once again that the probability distribution 
derived in this paper has some universal properties. 

From the probability distribution we are able to deduce the physical 
scattering amplitude. Again, we have considered the same interesting 
asymptotic behaviours as for the probability distribution: the vicinity 
of the average saturation scale, deep inside the saturation regime or 
in the dilute domain. Within this analysis, the most important result is 
that at high-energy the probability distribution can be considered as 
Gaussian over an very large domain $|z|\ll~\gamma_c\sigma^2~\propto~Y$. 
As a consequence the scattering amplitudes are dominated by events which 
are at saturation.

This validates the approach adopted in Refs. \cite{himst, ims}, and leads to the 
fact that scattering amplitudes scale as a function of $\log[r^2 \bar 
Q_s^2(Y)]/\sqrt{Y}$, 
a property which is known as {\em diffusive scaling} and may have important 
consequences on LHC physics. The domain over which the probability 
distribution 
may be considered as Gaussian extends over a region of $\rho_s$ which 
satisfies 
$|\rho_s-\avgrho|\ll \gamma_c\sigma^2$. As the dispersion $\sigma$ grows 
like 
$\sqrt{Y}$, this domain becomes larger and larger with increasing rpidity. The 
same 
argument holds for the region $|Z| \ll \gamma_c \sigma^2$ for which the 
scattering 
amplitude 
satisfies diffusive scaling. Those facts also corroborate the result from 
\cite{strong} that, in the strong-fluctuation limit, the scattering amplitude is 
given 
by the error function \eqref{eq:average_gauss}, which 
is a superposition of Heaviside functions with a Gaussian distribution.

Of course, the analytical behaviour of the cumulants associated with the 
distribution of $\rho_s$ for any value of $\alpha_s$ and in the case of the full 
QCD equation \eqref{eq:langevin} still remains an open question. However, our 
analytical and numerical results show  that they are compatible with a universal 
probability distribution. This deserves more detailed studies. 

\begin{acknowledgments}

We would like to thank St\'ephane Munier for his comments on the paper.
C.M. and B.X are grateful to Al Mueller for inspiring discussions which 
triggered this work. B.X. also thanks Arif Shoshi for discussions and 
comments. G.S. is funded by the National Funds for Scientific Research 
(Belgium).

\end{acknowledgments}

\end{document}